\documentclass[
journal=
apchd5%
,manuscript=
article%
]{achemso}

\usepackage[version=3]{mhchem} 
\usepackage[T1]{fontenc}       

\usepackage{graphicx}
\usepackage{amsmath}
\usepackage{comment}%
\usepackage[usenames,dvipsnames,svgnames,table]{xcolor}
\usepackage[colorlinks=true,linkcolor=blue]{hyperref}%

\SectionNumbersOff

\author{Jiun-Yi Lien}%
\affiliation{Department of Chemistry, National Tsing Hua University, Hsinchu 30013, Taiwan, Republic of China}
\altaffiliation{Contributed equally to this work}%
\author{Chih-Jung Chen}%
\affiliation{Nanomaterials Laboratory, Far East University, Hsing-Shih, Tainan 74448, Taiwan, Republic of China}
\altaffiliation{Contributed equally to this work}%
\author{Ray-Kuang Chiang}%
\email{rkc.chem@msa.hinet.net}%
\affiliation{Nanomaterials Laboratory, Far East University, Hsing-Shih, Tainan 74448, Taiwan, Republic of China}
\author{Sue-Lein Wang}%
\email{slwang@mx.nthu.edu.tw}%
\affiliation{Department of Chemistry, National Tsing Hua University, Hsinchu 30013, Taiwan, Republic of China}

\title[]{High color-rendering flat lamps using quantum-dot color conversion films}

\abbreviations{CQD, LED, FRET, CCT, LE, CRI, LER, EQE, QDCCF, ZnO, MMA, LMA, MA, AIBN, FWHM, PL, TEM, SAED, HRTEM, ZB, EDS}

\begin{document}

\begin{tocentry}
\centering
\includegraphics[scale=0.35]{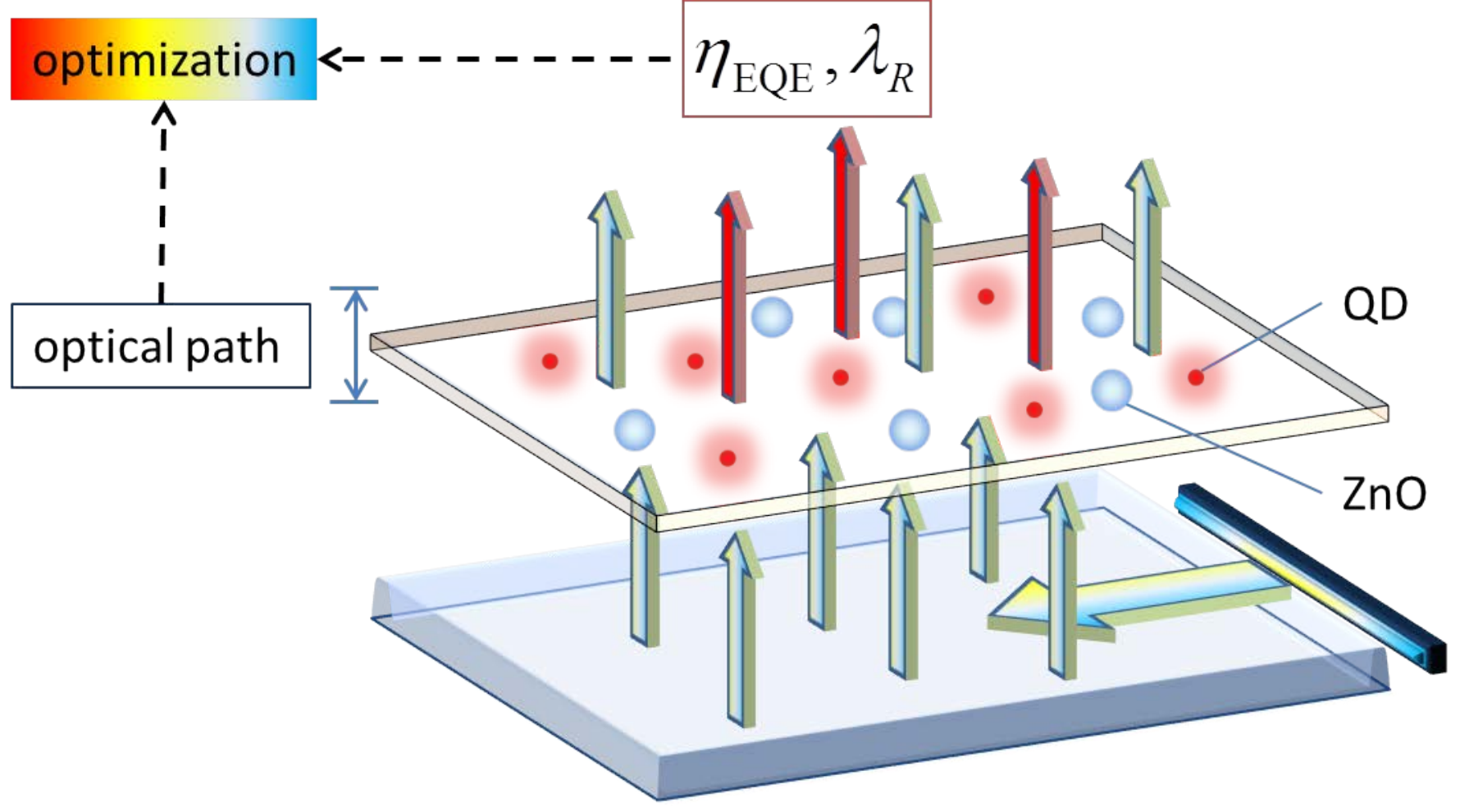}
\end{tocentry}

\begin{abstract}
Colloidal quantum dots are promising next-generation phosphors to enhance the color rendition of light-emitting devices (LEDs) while minimizing the brightness drop.
In order to exploit the beneficial tunability of quantum dots for highly efficient devices, optimization and determination of the performance limit are of crucial importance.
In this work, a facile preparation process of red-emission quantum dot films and simulation algorithm for fitting this film with two commercial LED flat lamps to the optimized performance are developed.
Based on the algorithm, one lamp improves from cold-white light (8669 K) with poor color rendition ($R_{a}=72$) and luminous efficacy (85 lm/W) to warm-white light (2867 K) with $R_{a}=90.8$ and $R_{9}=74.9$, and the other reaches $R_{a}=93\sim95$.
Impressively, the brightness drop is only about $15\sim20\%$.
Furthermore, our device shows reliability over 1000 hours with only PET (polyethylene-terephthalate) films as the barrier, indicating that this auxiliary red-emission film can be easily applied to improve the color rendition of most commercial LED flat lamps.
\end{abstract}

\section{Introduction}

Semiconducting nanocrystals, also called quantum dots (QDs), have attracted great attention due to promising applications such as light-emitting diodes (LEDs), solar cells, and biosensors \cite{Wood&Bulovic:NanoRev.1.5202(2010)(CQD_LEDs),Kim:Adv.Mater.25.4986(2013)(25th_Anniversary_CQDs),Shirasaki&Bawendi&Bulovic:NPhoton.7.13(2013)(Emergence_CQD_light-emitting_tech),Bera:Materials.3.2260(2010)(QDs&multimodal_applications_Review)}. 
For applications of lighting and displays, QDs can be utilized to enhance the color rendition of white-LED lamps \cite{Nizamoglu&Zengin&Demir:APL.92.031102(2008)(3-color_QDLED_warm-white_light),Nizamoglu&Erdem&Sun&Demir:Opt.Lett.35.3372(2010)(Warm-white_QDLED_4-colors)} and the color gamut of liquid-crystal displays \cite{Jang:Adv.Mater.22.3076(2010)(WLED_QD_BLU),Luo&Xu&WuST:J.Disp.Tech.10.526(2014)(Emerging_CQD_LCDs),Zhu&Luo&Chen&Dong&WuST:Opt.Exp.23.23680(2015)(Realizing_Rec.2020_C.G._QD_displays)}. 
Compared with traditional phosphors, the QDs are with advantages of high efficiency, tunable emission wavelength, narrow bandwidth, and low-cost solution-synthesis. 
However, with wide absorption spectra and nano-scale sizes, reabsorption and F\"orster resonance energy transfer (FRET) could induce the problems of color shift and efficiency drop, depending on the device structure \cite{Kagan&Murray&Nirmal&Bawendi:PRL.76.1517(1996)(FRET_CdSe_QD_solids),Jang_et_al.:Adv.Mater.20.2696(2008)(WLED_with_Excellent_CRI_CQD&Phosphors),Akselrod&Bulovic.et_al.:NanoLett.14.3556(2014)(Subdiffusive_exciton_transport_in_QD_solids)}.
Other issues like photo-oxidization \cite{Hines&Becker&Kamat:J.Phys.Chem.C.116.13452(2012)(Photo-oxidation&Exciton_dynamics_CdSe_QDs),Hines&Kamat:ACS.Appl.Mater.Interfaces.6.3041(2014)(Recent_Advances_QD_Surface_Chemistry)} and thermal quench \cite{Zhao&Meijerink:ACSNano.6.9058(2012)(High-Temperature_Luminescence_Quenching_CQDs),Cai&Kelley:J.Phys.Chem.C.117.7902(2013)(Thermal_Quenching_Mechanisms_II–VI_Semiconductor_NCs)} also limit QD applications. 
Many strategies have been adapted to avoid these issues, e.g., the remote-phosphor designs \cite{Moon&Yoo:Opt.Mater.Express.4.2174(2014)(Optical_characteristics_longevity_QD-coated_WLED)}.

In the field of solid-state lighting, it is crucial for indoor illumination to generate warm-white light with correlated color temperature (CCT) lower than 4000 K while maintaining high color rendition and high luminous efficacy (LE).
Theoretically, it can be generated by four-color LEDs to reach the color-rendering index (CRI) with $R_{a}=90$ and the highest luminous efficacy of radiation (LER) being 408 lm/W$_\textrm{opt}$, provided that the intensity ratios and wavelengths of the four colors are given correctly \cite{Phillips:Laser&Photon.Rev.1.307(2007)(Research_challenges_ultra-efficient_inorganic_SSL)}.
Due to the green gap of LED technology, high cost of additional control circuits, and different lifetimes of individual chips, the devices using multiple chips are inferior to the ones using a blue-emitting chip together with multiple phosphors \cite{Krames:J.Disp.Tech.3.160(2007)(Status_future_high-power_LED_for_SSL)}.
It was predicted that, by employing QD phosphors, a LED has the potential to achieve LER larger than 380 lm/W$_\textrm{opt}$ and LE being 256 lm/W with perfect light recycle and extraction efficiency \cite{Erdem&Nizamoglu&Sun&Demir:Opt.Express.18.340(2010)(photometric_investigation_LEDs_with_high_CRI&LE_employing_NQDs),Zhong&He&Zhang:Opt.Express.20.9122(2012)(Optimal_spectra_QD-WLED),Erdem&Nizamoglu&Demir:Opt.Express.30.3275(2012)(Computational_study_PCE&LE_QDLED)}.

Although the electrical-to-optical power conversion efficiency of commercial blue-emitting chips is improved every year, the LE of current lighting devices (using traditional phosphors, QDs, or both) is still far from the theoretical limit to date \cite{Song&Lee&Yang:Opt.Mater.Express.3.1468(2013)(Fabrication_warm_high_CRI_WLED_non-cadmium_QDs),Kim&Song&Yang:Opt.Lett.38.2885(2013)(Color-converting_bilayered_composite_plate_QD-polymer_high-CRI_WLED),Moon&Yoo:Opt.Mater.Express.4.2174(2014)(Optical_characteristics_longevity_QD-coated_WLED)}.
Besides the imperfect down-conversion and light-extraction efficiencies of phosphors and QDs, multiple-color mixing also causes the problems of reabsorption and FRET \cite{Jang_et_al.:Adv.Mater.20.2696(2008)(WLED_with_Excellent_CRI_CQD&Phosphors),Demir:NanoToday.6.632(2011)(QD-LED_using_photonic_&_excitonic_color_conversion),ChenCJ&LinCC&LienJY&WangSL&ChiangRK:JMCC.3.196(2015)(QD-polymer_color_mixing_with_alleviated_FRET),Lee_et_al.:ACSNano.5b05513(2015)(RGB_QD-Mixed_Multilayer_QLED)}.
In order to simultaneously meet the demands of CCT, CRI, and LE, optimization of the wavelength, bandwidth, and quantity of QDs has to be reached for various optical structures and phosphor compositions with limited efficiencies.
However, to our best knowledge, there is still no such an optimization method available in the literature.

In this work, we develop an algorithm for optimization by determining the optical path of QD absorption and the external quantum efficiency (EQE) of QD emission.
We also produce the desired quantum-dot color-conversion films (QDCCFs) and apply them to  
two commercial flat lamps of cold-white light (CCT $=8669$ K, labeled as lamp-1) and day-white light (CCT $=5792$ K, labeled as lamp-2).
By employing a simple film-type structure, we achieve warm-white light (CCT $=2867$ K) with $R_{a}>90$, $R_{9}>70$ for the strong red color, and \textit{wall-plug} LE $=68$ lm/W, 
better than most reported phosphor-based technologies \cite{Shirasaki&Bawendi&Bulovic:NPhoton.7.13(2013)(Emergence_CQD_light-emitting_tech)}.
More importantly, the drop of brightness due to lowering the CCT is limited to only $15\sim20\%$ without modification of the basic structure of the lamps.
Our device also shows excellent stability and can be directly applied to most of commercial LED flat lamps.

\section{Results and discussion}

\subsection{Well-dispersed QD-polymer film}


\begin{figure}
  \includegraphics[width=0.49\linewidth]{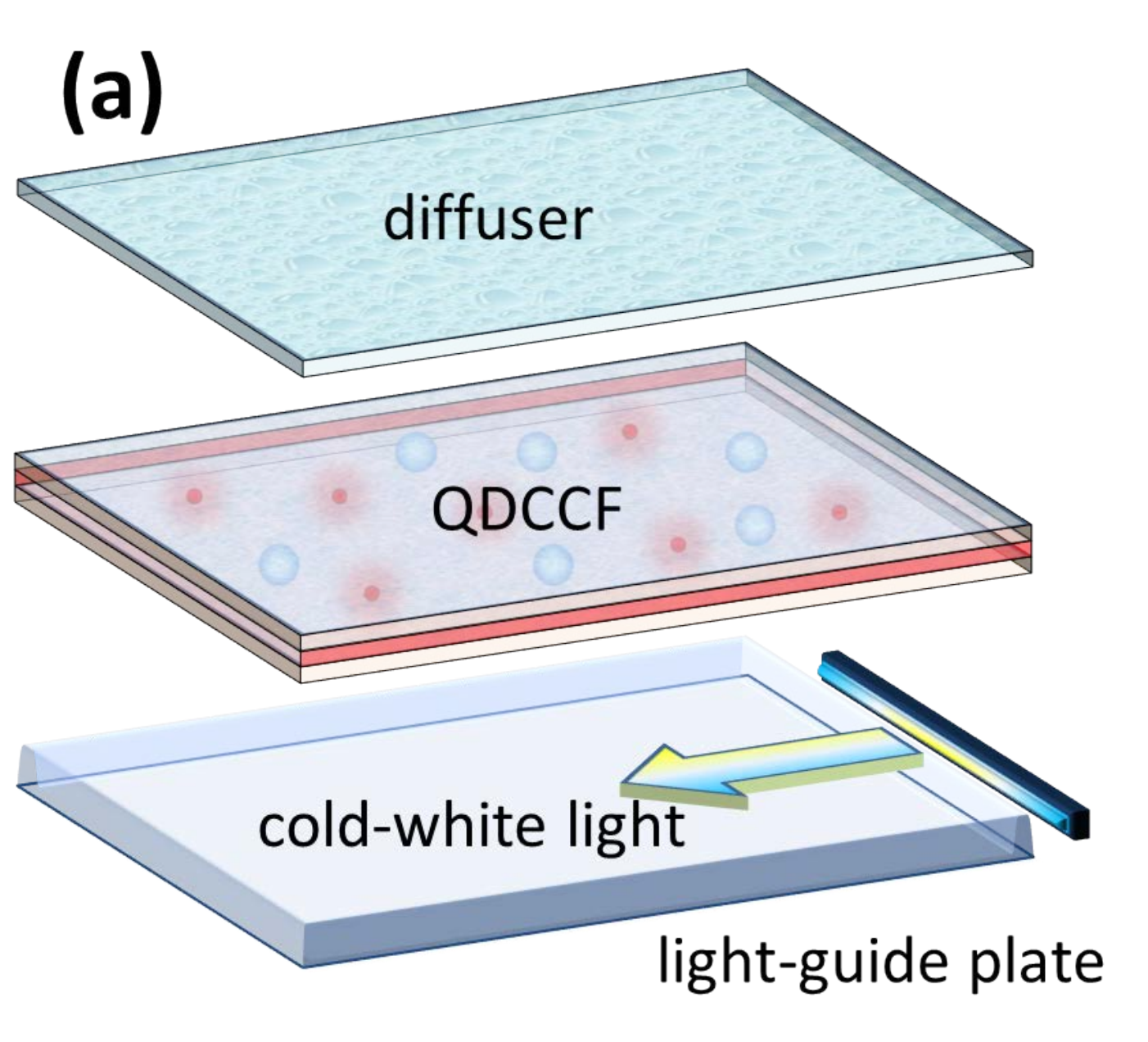}
  \includegraphics[width=0.45\linewidth]{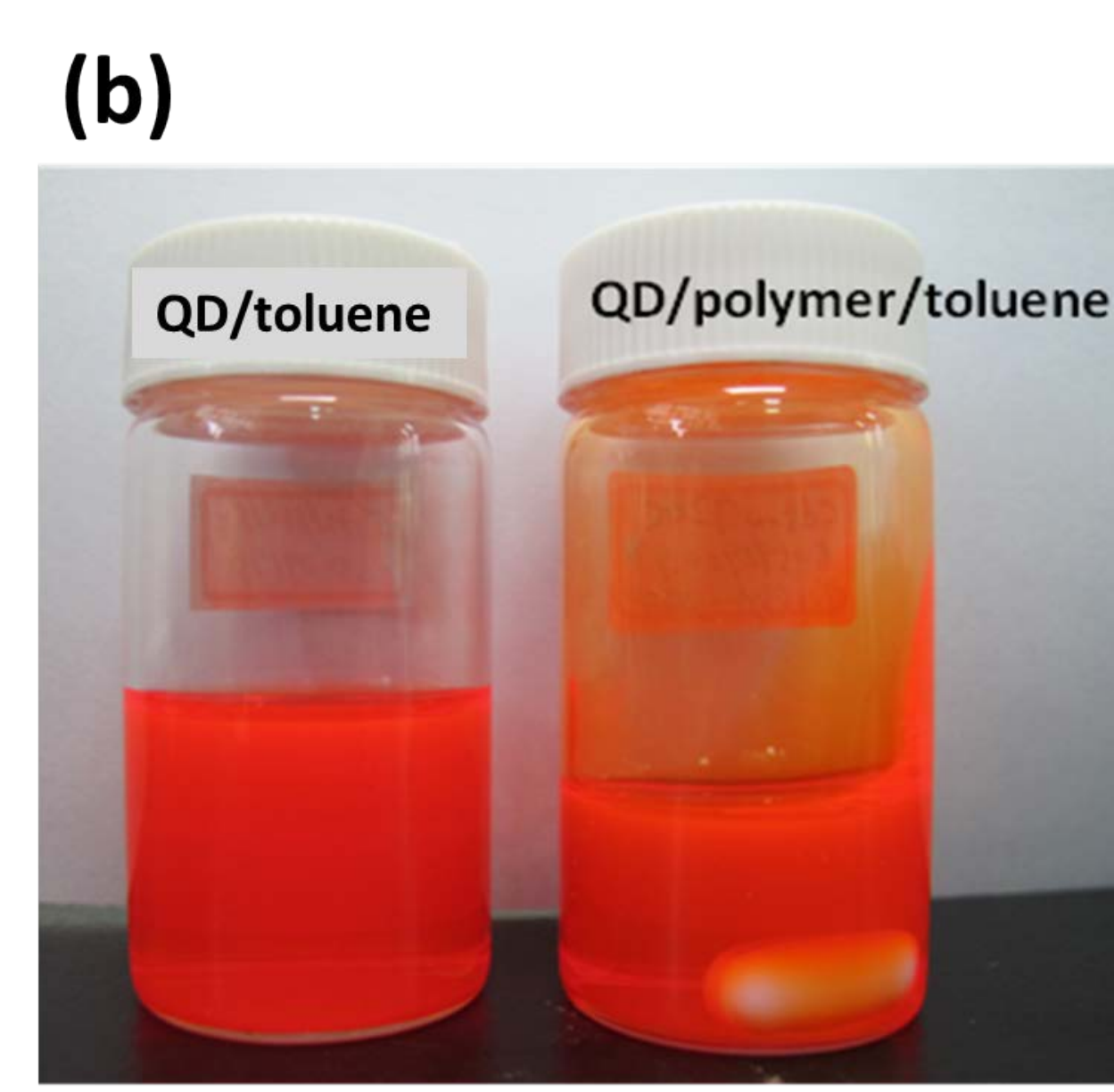}
  \includegraphics[width=0.95\linewidth]{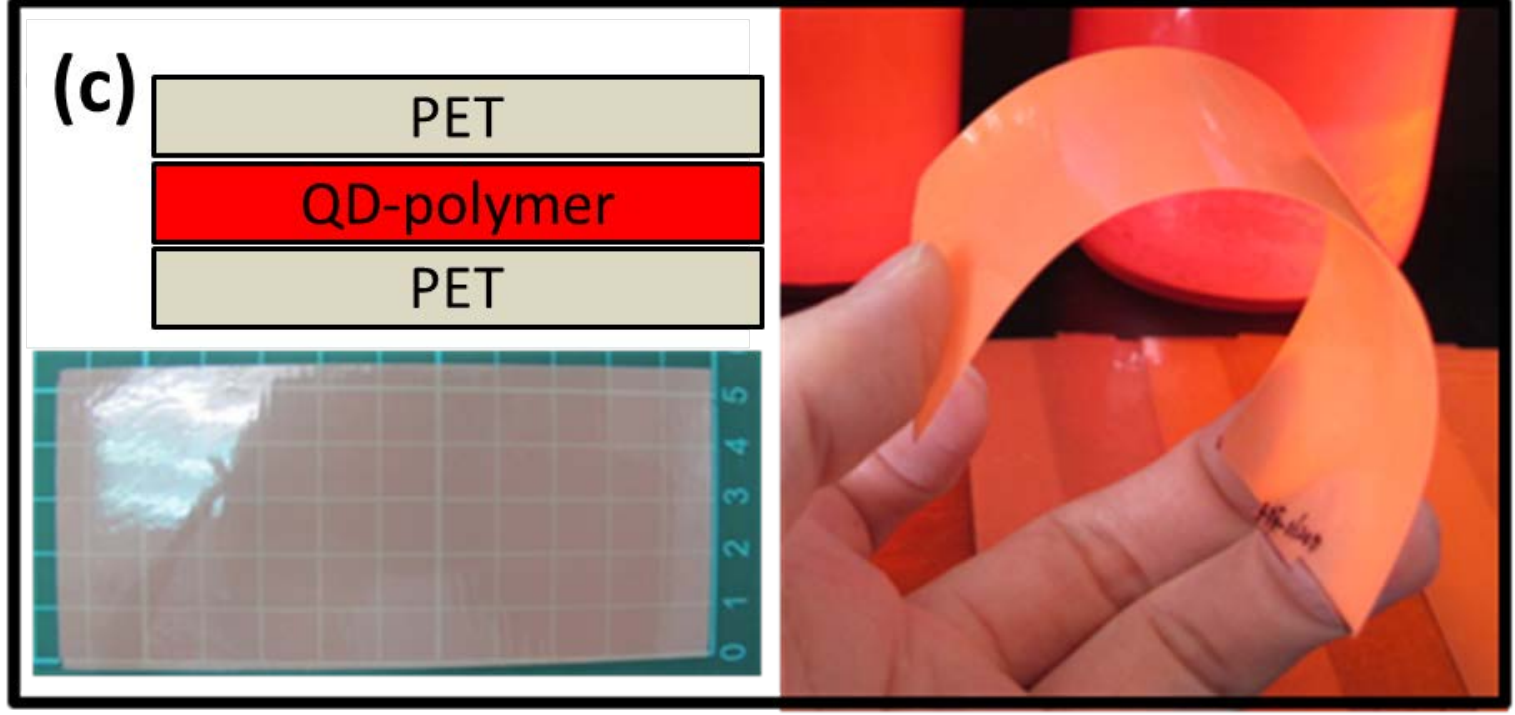}
  \caption{
  (a) Device structure of the warm-white flat lamps employing the QDCCFs.
  (b) Photographs of well dispersed QDs in toluene and toluene-polymer solutions. 
  (c) Demonstration of the structure and flexibility of the QDCCFs.
  }\label{QDCCF_characteristics}
\end{figure}


The commercial flat lamps used in our experiment consist of a traditional edge-light structure with a diffuser film to enhance the uniformity of light, 
as shown in Fig.~\ref{QDCCF_characteristics}(a).
For the warm-white light application, the QDCCFs are simply inserted between the light-guide plate and diffuser film.
The QDCCFs were fabricated by creating a homogeneously dispersed QD-polymer composite [Fig.~\ref{QDCCF_characteristics}(b)], where the functionalized co-polymer containing a small amount of methacrylic acid (1\%) was designed to increase the affinity to QD surfaces \cite{Tamborra_et_al.:Small.3.822(2007)(QD-PMMA_Composites_for_Nanoimprinting_Lithography)} and QDs are tightly held on the polymer chains for dispersion \cite{ChenCJ&LinCC&LienJY&WangSL&ChiangRK:JMCC.3.196(2015)(QD-polymer_color_mixing_with_alleviated_FRET)}.
The QD-polymer composite is sealed by two polyethylene-terephthalate (PET) films to reduce photo-oxidation, and the flexibility of QDCCFs is improved by adding 24 wt\% Lauryl methacrylate (LMA) in the co-polymer, 
as shown in Fig.~\ref{QDCCF_characteristics}(c). 
In addition, the zinc-oxide (ZnO) scattering particles are dispersed inside the QD-polymer layer in order to enhance the light-extraction efficiency from the guiding modes \cite{Ryu&Hong_et_al.:JAP.109.093116(2011)(Enhanced_extraction_GaN_LED_ZnO_rough_ITO), Yang&Demir_et_al.:Adv.Funct.Mater.24.5977(2014)(ZnO_nanopillar_array_enhanced_extraction_QLED), Mao_et_al.:ACS.Appl.Mater.Interfaces.7.19179(2015)(Polystyrene-ZnOMicronano_Hierarchical_Structure_Light_Extraction_LED), Shin_et_al.:Opt.Express.23.A133(2015)(ZnO_nanoparticle_layer_improving_extraction_OLED)}.

\begin{figure}
  \includegraphics[width=0.9\linewidth]{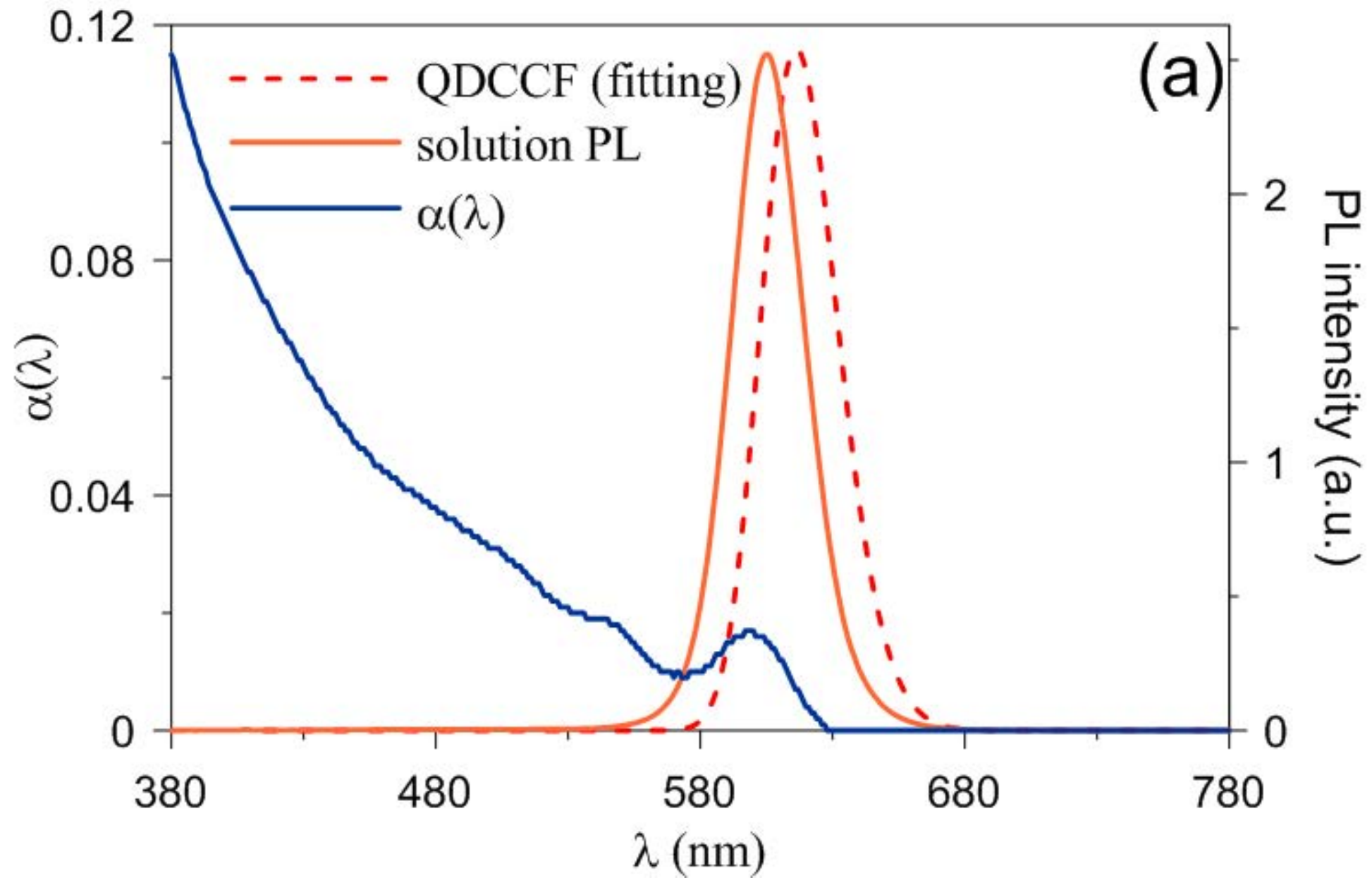}
  \includegraphics[width=0.9\linewidth]{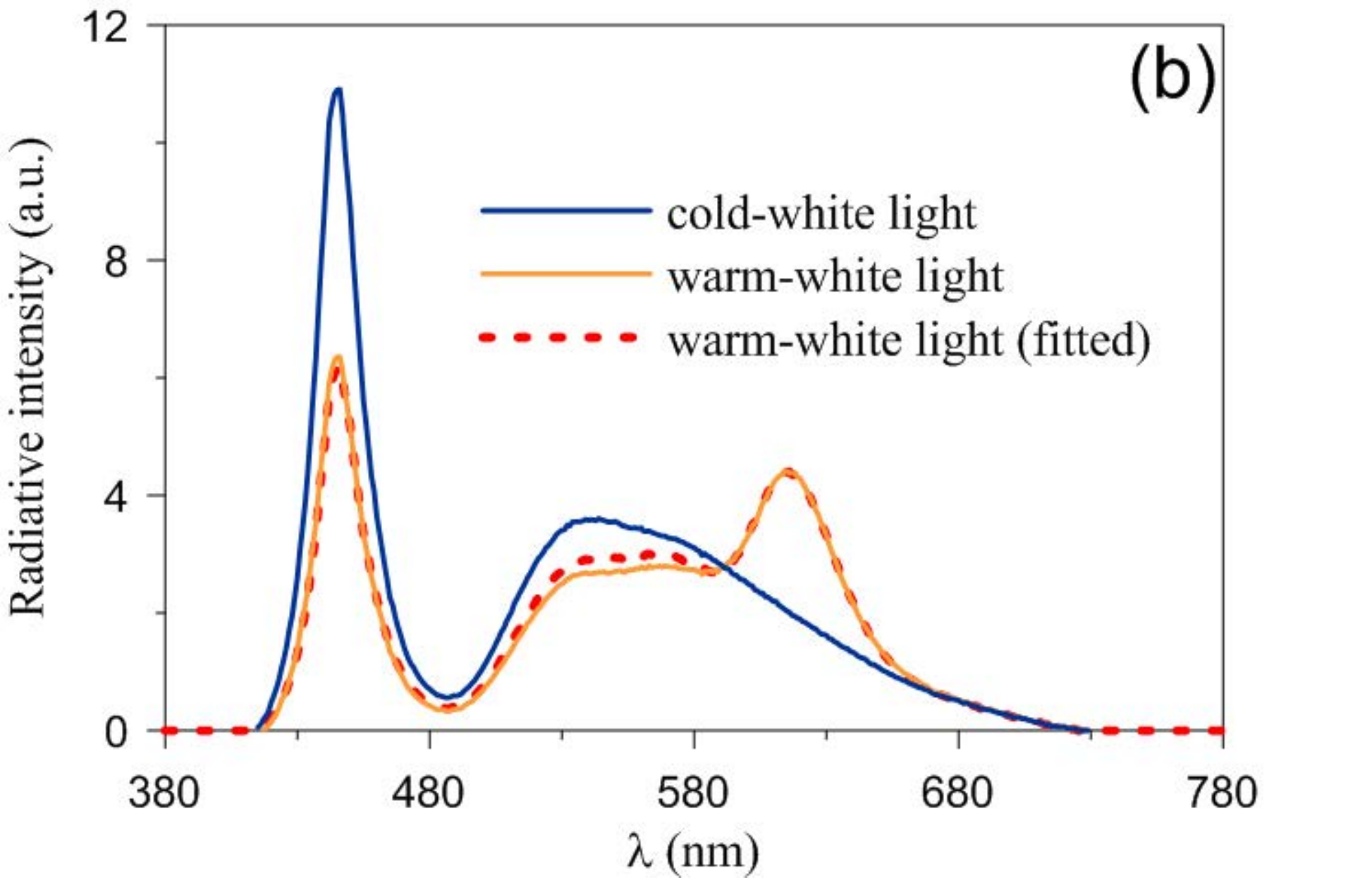}
  \caption{
  (a) The measured attenuation coefficient $\alpha(\lambda)$ and photoluminescent intensity of a dilute QD solution for the test sample (blue and orange solid curves), as well as the fitted red emission of the test QDCCF (dashed curve) with red shift about 6 nm.
  (b) The energy spectra of the input cold-white light (blue solid curve), the measured warm-white light of the test sample (orange solid curve), and the fitted warm-white light derived from our simulation algorithm (dashed red curve).
  }\label{test-sample}
\end{figure}

\subsection{Optimization algorithm}

For demonstration of our optimization scheme, the QDCCFs are applied to lamp-1 of cold-white light (CCT $=8669$ K) with poor colorimetric ($R_{a}=72$) and photometric (LE $=85$ lm/W) performance.
The scheme is based on the color-conversion process of QDCCFs, where partial cold-white light is absorbed by QDs and down-converted into red emission.
According to the Beer-Lambert law, the transmission spectrum $T_\textnormal{cw}(\lambda)$ of the cold-white light with wavelength $\lambda$ satisfies
\begin{align}
  T_\textnormal{cw}(\lambda)
  &=
  10^{-A(\lambda)}
  =
  10^{-\alpha(\lambda)\rho l(\lambda)}
  \,,\label{Beer-Lambert law}
\end{align}
where the absorbance $A(\lambda)$ is proportional to the attenuation coefficient $\alpha(\lambda)$ and the concentration $\rho$ of the QDs, as well as the optical path $l(\lambda)$ inside the device.
The absorbed photon number $N_\textnormal{abs}$ is determined by
\begin{align}
  N_\textnormal{abs}
  &=
  \int
  \left[
  1
  -
  T_\textnormal{cw}(\lambda)
  \right]
  n_\textnormal{cw}(\lambda)
  d\lambda
  \,,
\end{align}
where $n_\textnormal{cw}(\lambda)$ is the photon-density spectrum of the cold-white light.
Therefore, the down-converted photon number $N_\textnormal{emit}$ is given by
\begin{align}
  N_\textnormal{emit}
  &=
  \eta_\textnormal{EQE}
  \times
  N_\textnormal{abs}
  =
  \eta_\textnormal{QY}
  \times
  \eta_\textnormal{ext}
  \times
  N_\textnormal{abs}
  \nonumber\\
  &=
  \int
  n_\textnormal{r}(\lambda)
  d\lambda
  \,,
\end{align}
where $n_\textnormal{r}(\lambda)$ is the photon-density spectrum of the red emission, $\eta_\textnormal{EQE}$ is the external quantum efficiency of QDCCFs, $\eta_\textnormal{QY}$ is the photoluminescent quantum yield of QDs, and $\eta_\textnormal{ext}$ is the \textit{effective} light-extraction efficiency of the films.
The energy spectrum of the warm-white light $E_\textnormal{ww}(\lambda)$ is therefore determined by both the spectra of transmitted cold-white light $[1-T_\textnormal{cw}(\lambda)]E_\textnormal{cw}(\lambda)$ and red emission $E_\textnormal{r}(\lambda)$, i.e.,
\begin{gather}
  E_\textnormal{ww}(\lambda)
  =
  \left[
  1
  -
  T_\textnormal{cw}(\lambda)
  \right]
  E_\textnormal{cw}(\lambda)
  +
  E_\textnormal{r}(\lambda)
  \nonumber\\
  \equiv
  \left\{
  \left[
  1
  -
  T_\textnormal{cw}(\lambda)
  \right]
  n_\textnormal{cw}(\lambda)
  +
  n_\textnormal{r}(\lambda)
  \right\}
  \left(
  {hc\over\lambda}
  \right)
  \,,\label{down-converted energy spectrum}
\end{gather}
where $h$ is the Plank constant and $c$ is the speed of light.

In order to achieve maximum colorimetric or photometric performance with a given CCT range, we have to determine the appropriate absorbance $A(\lambda)$ and red emission $n_\textnormal{r}(\lambda)$ of the QDCCFs.
For $A(\lambda)$, the attenuation coefficient $\alpha(\lambda)$ can be determined by the absorbance of a dilute solution of the QDs, and the concentration $\rho$ is predetermined before fabricating the QDCCFs.
The optical path length $l(\lambda)$ depends on the film thickness, the density of scattering particles, the recycle mechanism of the device, etc.
Ignoring the small dispersion of the refractive indices of the device inside the optical regime, $l$ is approximately independent of the wavelength of the cold-white light and to be determined.

The red-emission spectrum $n_\textnormal{r}(\lambda)$ 
of the warm-white device cannot be determined by the photoluminescence spectra of either the QDCCFs or solutions due to a strong red shift \cite{ChenCJ&LinCC&LienJY&WangSL&ChiangRK:JMCC.3.196(2015)(QD-polymer_color_mixing_with_alleviated_FRET)}. 
This red shift results from FRET and guiding-induced reabsorption inside a film, and it is further enhanced by the recycle mechanism of the light-guide plate.
However, we can simulate $n_\textnormal{r}(\lambda)$ by a skew-normal distribution $f_\textnormal{sn}(\lambda)$, i.e.,
\begin{align}
  n_\textnormal{r}(\lambda)
  &=
  \eta_\textnormal{EQE}
  \times
  N_\textnormal{abs}
  \times
  f_\textnormal{sn}(\lambda)
  \\
  f_\textnormal{sn}(\lambda)
  &=
  {2\over\Delta_\textnormal{r}}
  \phi\left({\lambda-\lambda_\textnormal{r}\over\Delta_\textnormal{r}}\right)
  \Phi\left(\gamma\left({\lambda-\lambda_\textnormal{r}\over\Delta_\textnormal{r}}\right)\right)
  \,,\nonumber
\end{align}
where $\phi(x)$ is the standard normal probability density function and $\Phi(x)$ is the corresponding cumulative distribution function.
Here $\gamma$ is a parameter controlling the skewness of the distribution.
The parameters $\lambda_\textnormal{r}$ and $\Delta_\textnormal{r}$ are related (not identical, due to the skewness) to  the peak wavelength ($\lambda_\textnormal{R}$) and the full width at half maximum (FWHM, $\Delta_\textnormal{R}$) of the red emission.
The approximation of skew-normal distribution describes the band tail of a solid-state phosphor, and it can be generally applied to monodispersed QDs and many phosphors with a single emission level.

\begin{figure*}
  \includegraphics[width=0.465\linewidth]{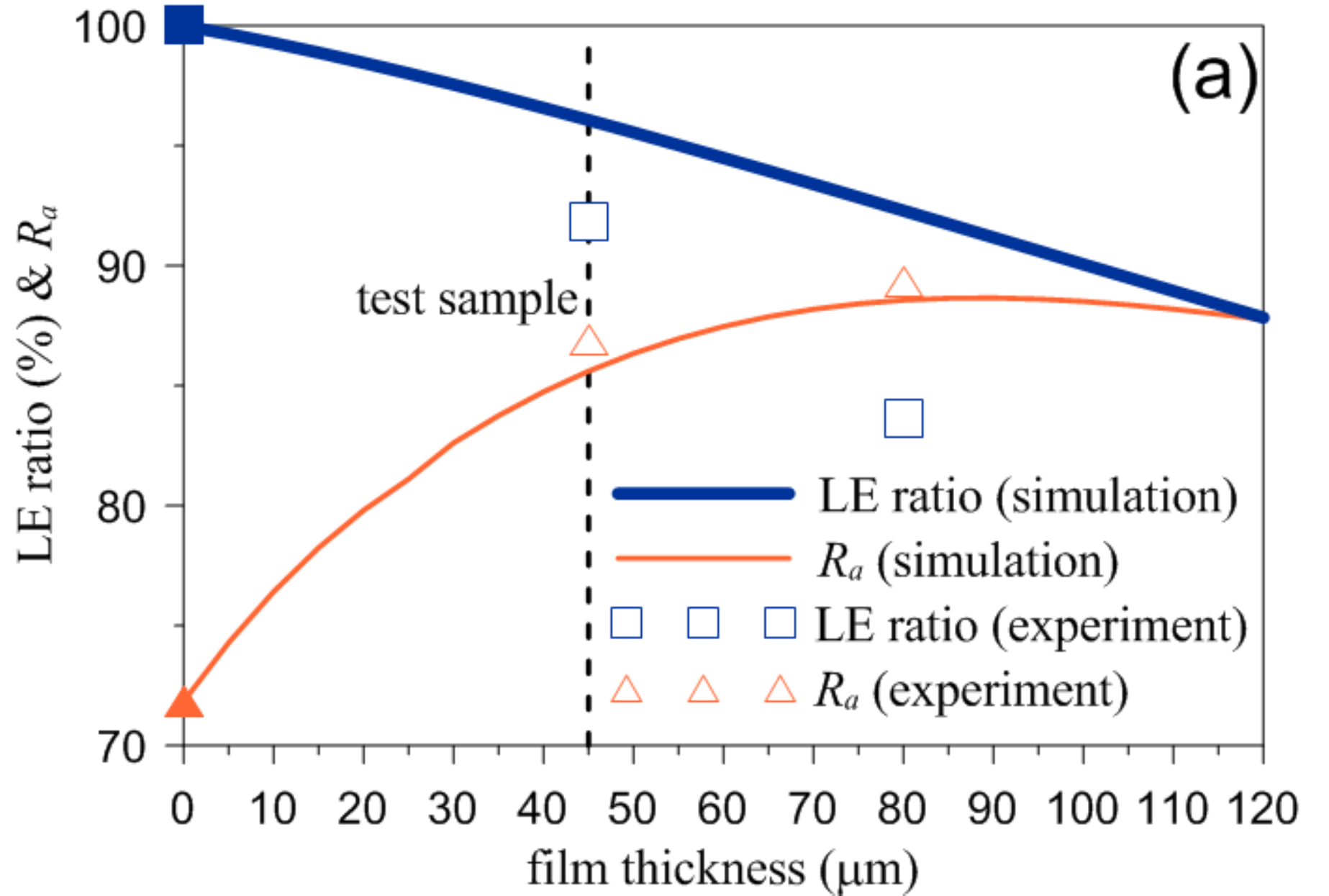}
  \includegraphics[width=0.52\linewidth]{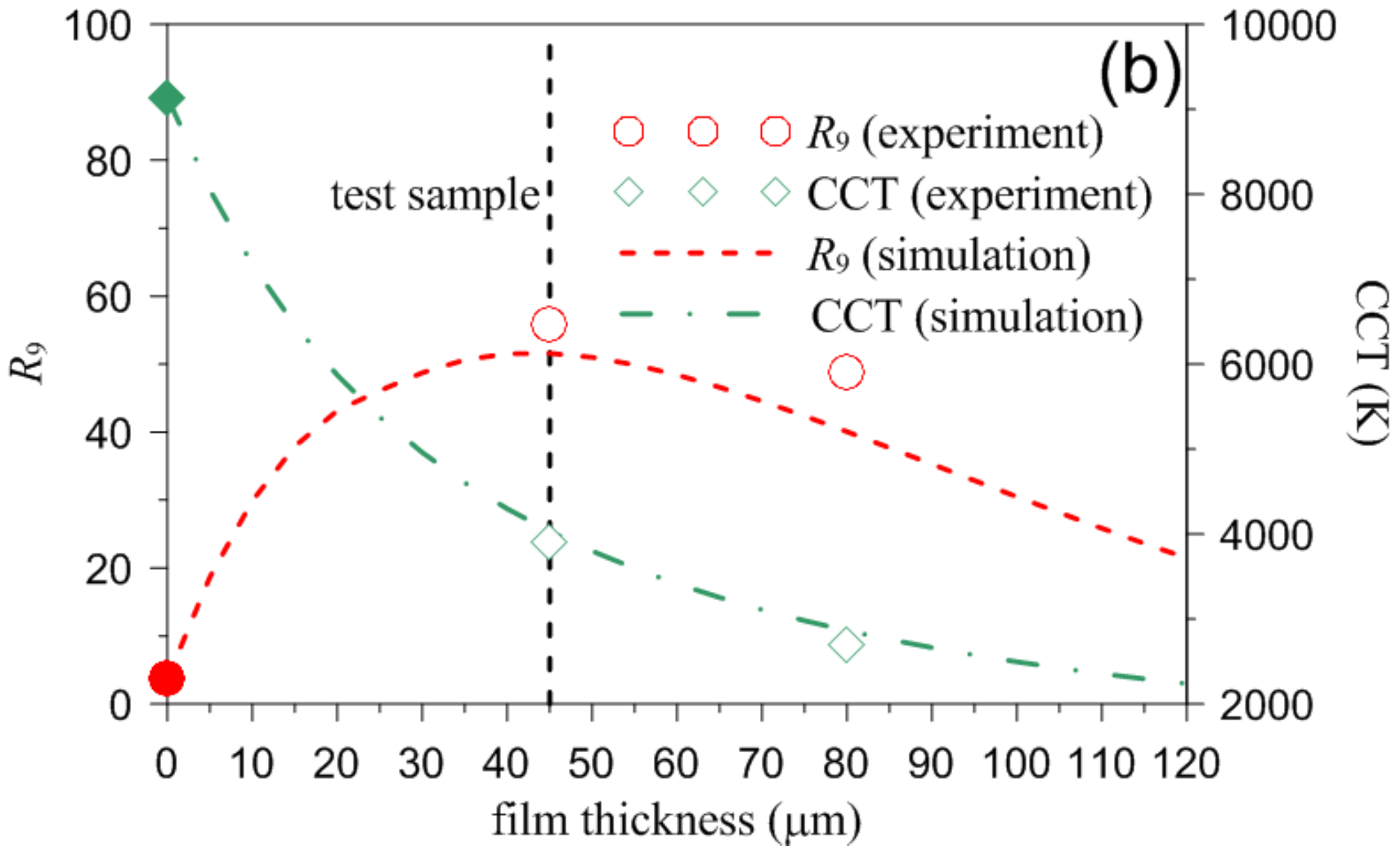}
  \includegraphics[width=0.465\linewidth]{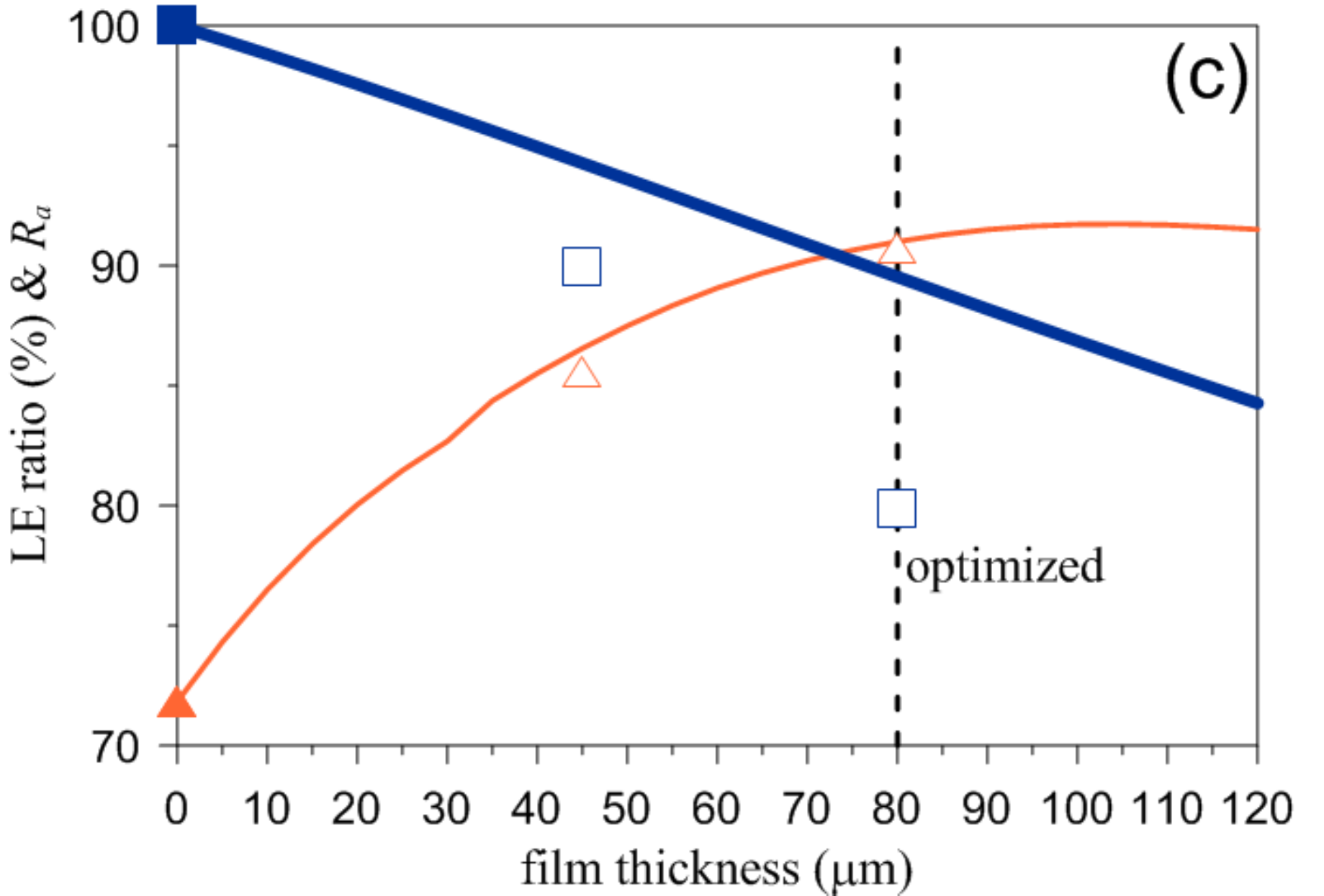}
  \includegraphics[width=0.52\linewidth]{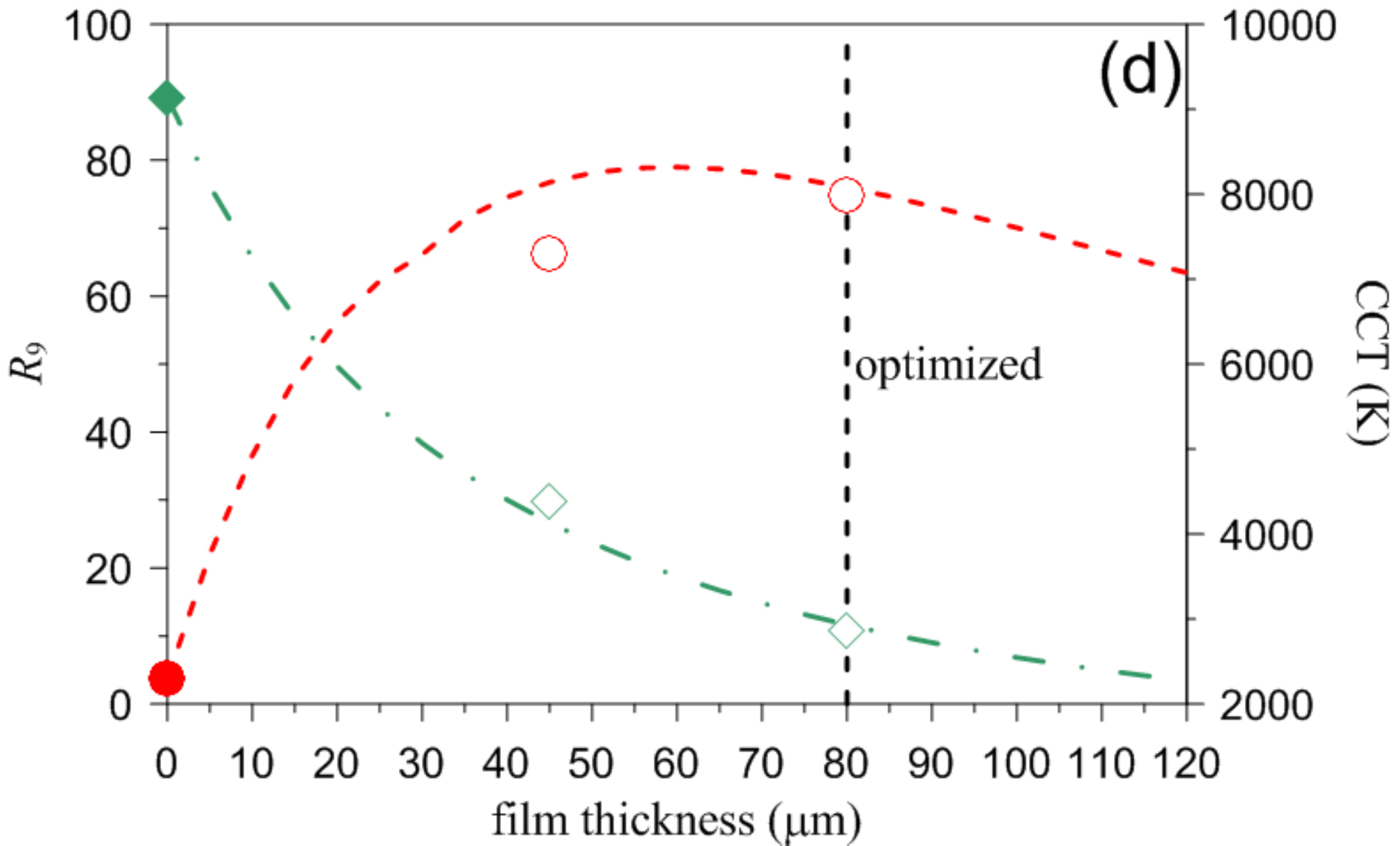}
  \includegraphics[width=0.465\linewidth]{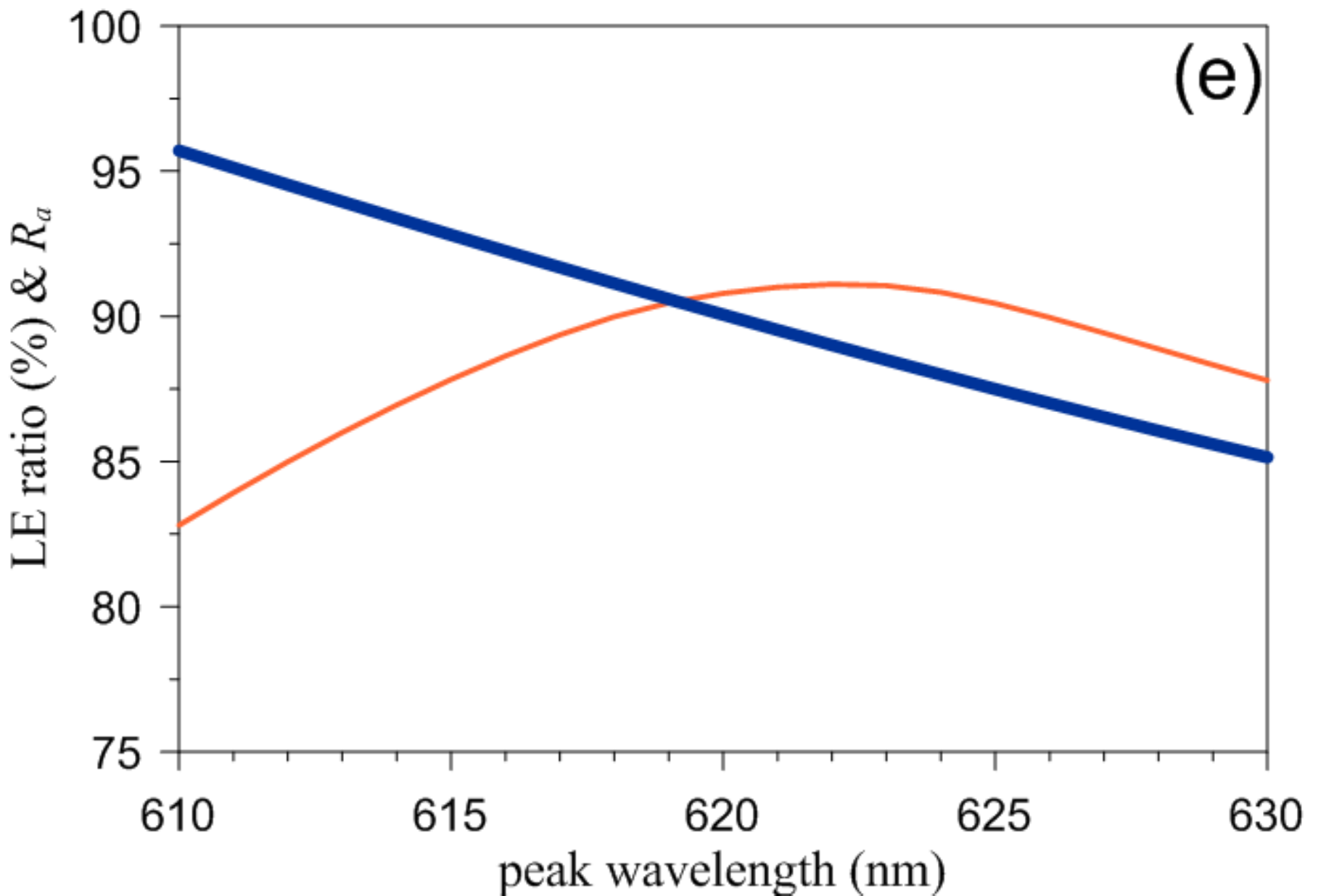}
  \includegraphics[width=0.52\linewidth]{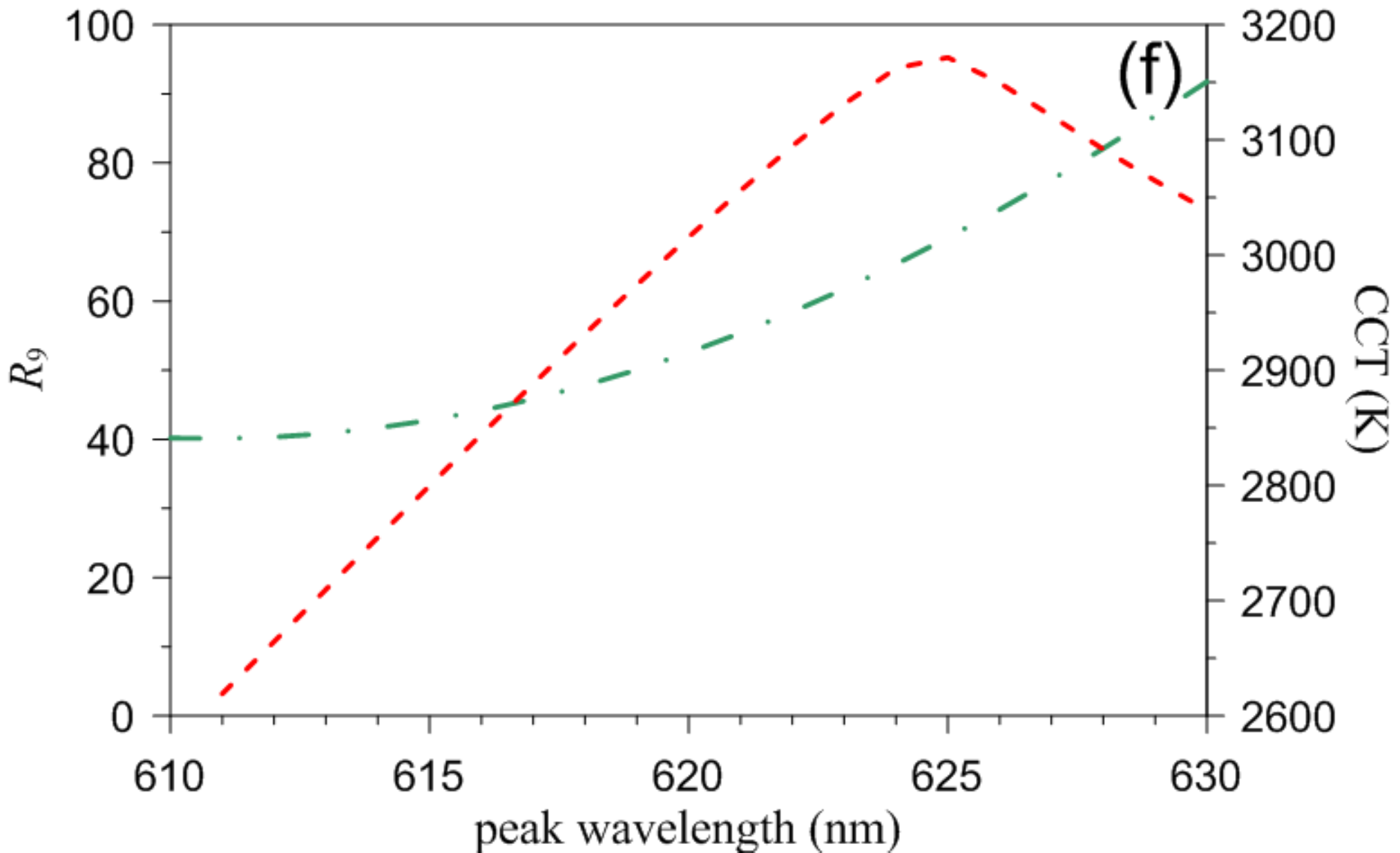}
  \caption{
  (a)-(d) Simulation of photometric (LE ratio with respect to the input cold-white light) and colorimetric performance as functions of the film thickness of the QDCCFs, where the wavelength ($\lambda_\textnormal{R}$) of red emission is 616 nm for (a)-(b) and 621 nm for (c)-(d), respectively. 
  (e)-(f) The same simulation as functions of $\lambda_\textnormal{R}$, with the film thickness being fixed at 80 $\mu$m.
  For comparison, the experimental data are also shown by discrete symbols. 
  The data of input cold-white light are represented as filled symbols, while those of the test sample and other samples for verification are represented as empty symbols.
  }\label{thickness & wavelength simulation}
\end{figure*}

The unknown parameters $l$, $\gamma$, $\lambda_\textnormal{r}$, $\Delta_\textnormal{r}$, and $\eta_\textnormal{EQE}$ can be determined by fitting the energy spectrum $E_\textnormal{ww}(\lambda)$ of a test QDCCF, as shown in Fig.~\ref{test-sample}.
The thickness of the test QDCCF (QD-polymer layer) is about 45 $\mu$m, and the emission wavelength/FWHM ($\lambda_\textnormal{R}$/$\Delta_\textnormal{R}$) given by $\gamma$, $\lambda_\textnormal{r}$, and $\Delta_\textnormal{r}$ is 616/34 nm, which is red-shifted about 6 nm from that of the same QDs inside a dilute solution [see Fig.~\ref{test-sample} (a)].
The external quantum efficiency $\eta_\textnormal{EQE}$ of the test sample is about 69\%.
For the QDs of the test sample, the quantum yield ($\eta_\textnormal{QY}$) is about 90\%, and the effective light-extraction efficiency $\eta_\textnormal{ext}$ can be estimated as 77\%.
The optical path length $l$ is proportional to the film thickness with fixed QD density and device structure, and thus we can find an optimized thickness by tuning $l$.
Another way to control the absorbance $A(\lambda)$ is to tune the QD density $\rho$.
However, increasing $\rho$ leads to a further red shift and the reduction of $\eta_\textnormal{EQE}$ due to reabsorption and FRET.
Thus we optimize the QDCCF with a fixed $\rho$ of 1 wt$\%$ for simplicity.

From Fig.~\ref{test-sample} (b), 
the only deviation of the measured spectrum is the slightly reduced yellow emission from LED phosphors, compared with the calculated spectrum.
This is because of that a small part of the blue light recycled by the light-guide plate and diffuser film to excite the yellow emission is absorbed by the QDCCF.
Although this effect leads to an overestimation of LE, we will see later that all the trends of photometric and colorimetric performance can be well predicted and used to optimize the QDCCF.

\subsection{Optimized warm-white-light devices}

The simulated properties with various film thickness and emission wavelength $\lambda_\textnormal{R}$ of the QDCCFs are shown in Fig.~\ref{thickness & wavelength simulation}, based on the parameters derived from the test sample of warm-white light and the input cold-white light.
In general, both LE and CCT decrease with increasing film thickness [Fig.~\ref{thickness & wavelength simulation} (a)-(d)]. 
The CRIs increase with film thickness to their maximum values, and then drop slowly due to deviation from the reference black-body radiation.
Compared with the experiments, all colorimetric properties are well predicted by our simulation.
The overestimation of LEs increases with film thickness due to that more recycled blue light is shared with the QDCCF.
However, the resulting dropping trends of LE simulation are consistent with the experimental ones, and is sufficient to determine the optimal thickness.

Our goal is to achieve $R_{a}\geq90$ and $R_{9}\geq70$ with the largest LE.
Using the same QDs of 616 nm emission as the test sample, the maximum $R_{a}$ is about $88.7$ with thickness 90 $\mu$m, and thus the emission wavelength of the QDCCF must be further tuned.
From the simulation of wavelength effects in Fig.~\ref{thickness & wavelength simulation} (e)-(f), 
the LE decreases with increasing $\lambda_\textnormal{R}$ because of the contribution of luminosity function of human visual perception.
The CCT slightly increases with $\lambda_\textnormal{R}$ and can be controlled around 3000 K with film thickness $70\sim80~\mu$m.
The general color rendering index $R_{a}$ reaches 90 only for $\lambda_\textnormal{R}$ in the range of $619\sim625$ nm.
Moreover, the red-color rendering index $R_{9}$ can be significantly enhanced to above 70 for $\lambda_\textnormal{R}=621\sim630$ nm, and it even reaches 95 for $\lambda_\textnormal{R}=625$ nm.
Therefore, we prepared a QDCCF with the desired wavelength about 621 nm with film thickness 80 $\mu$m, 
which is already shown in Fig.~\ref{thickness & wavelength simulation} (c)-(d).
This optimized sample exhibits $R_{a}=90.8$, $R_{9}=74.9$, CCT $=$ 2867 K, and LE $=$ 68 lm/W 
($80\%$ brightness compared with the cold-white light of lamp-1).

\begin{figure}
  ~~
  \includegraphics[width=0.5\linewidth]{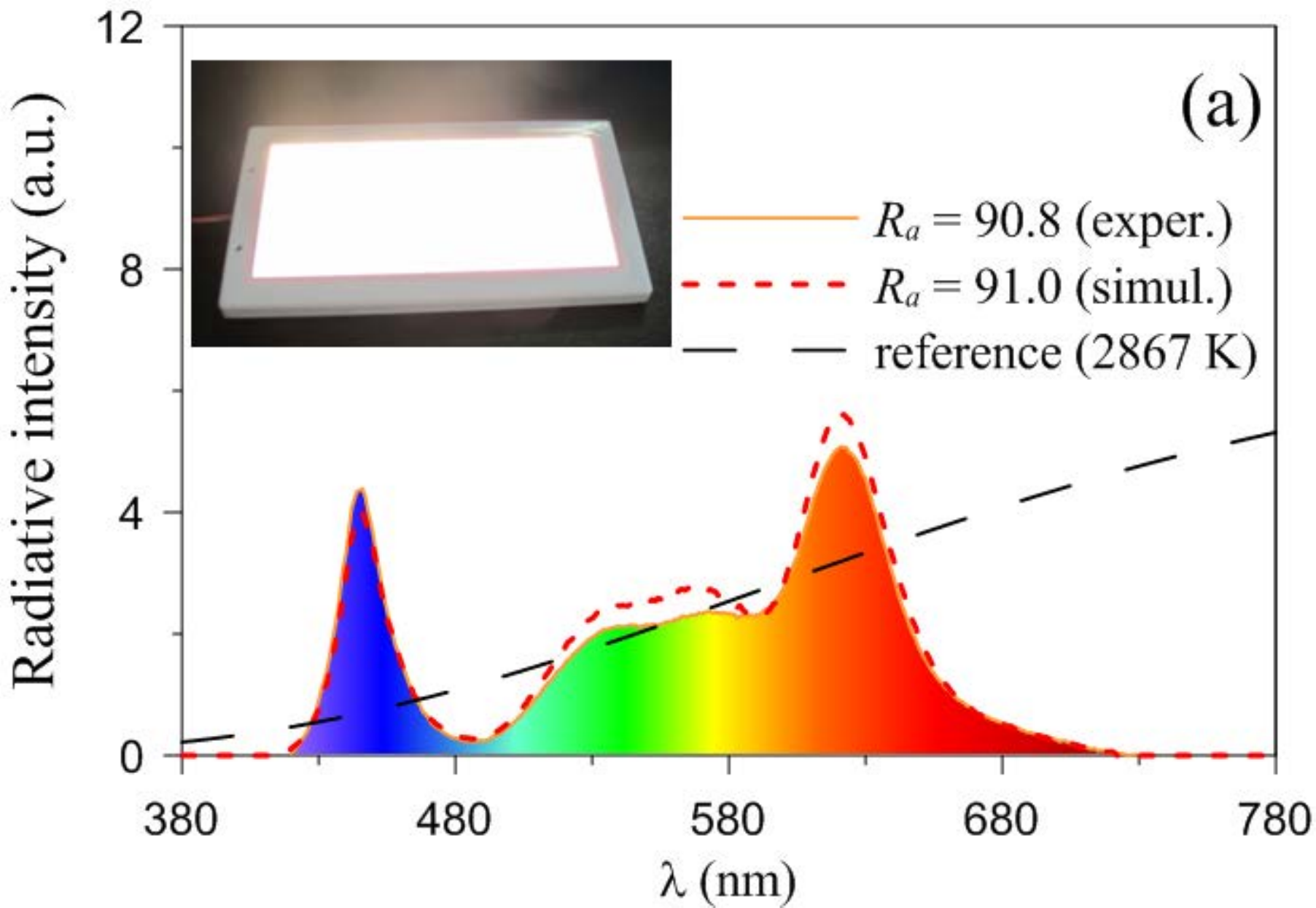}
  \\
  ~~
  \includegraphics[width=0.5\linewidth]{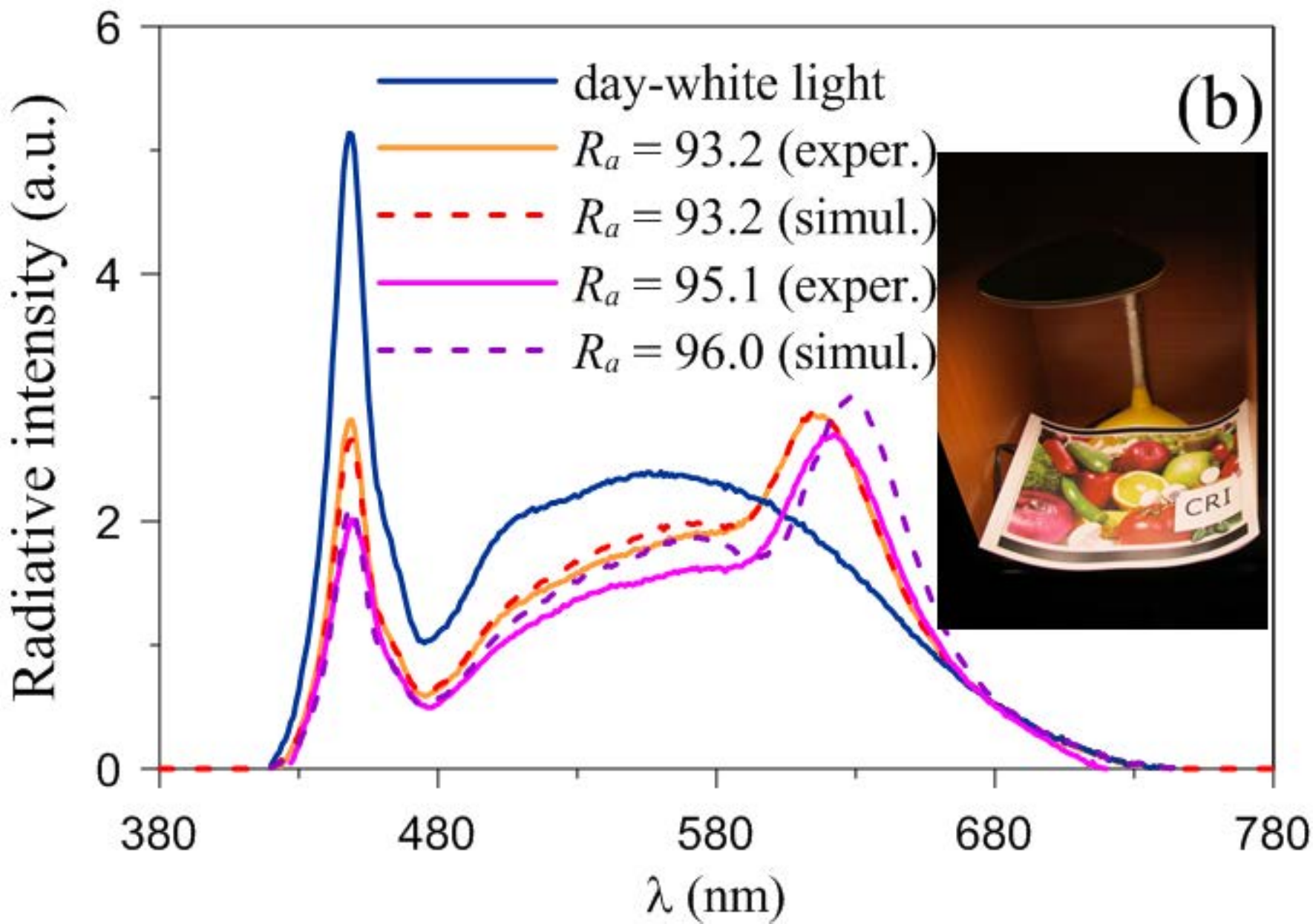}
  \includegraphics[width=0.5\linewidth]{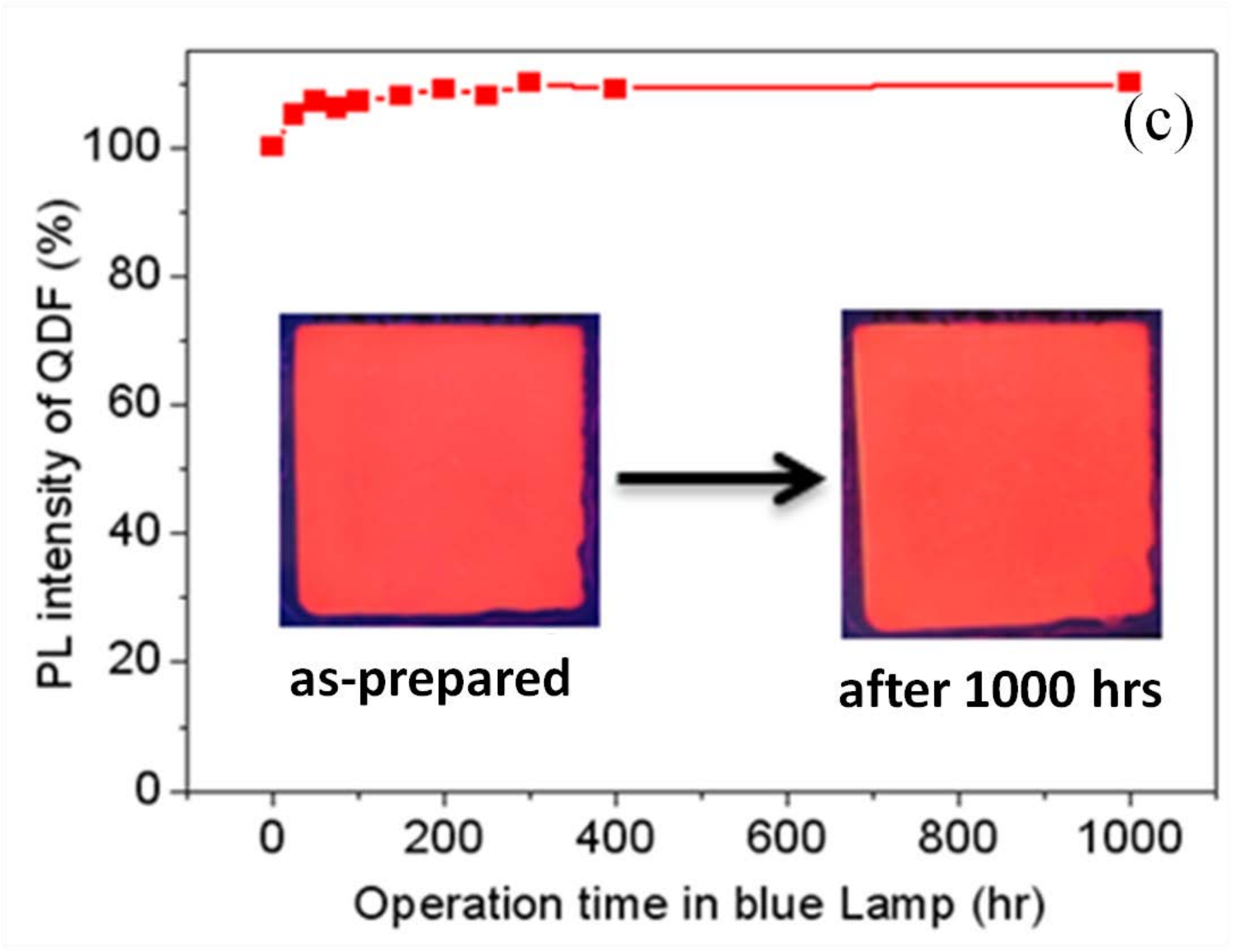}
  \caption{
  (a) Measured energy spectrum (orange solid curve) of the optimized warm-white light of lamp-1 with $R_{a}=90.8$, $R_{9}=74.9$, CCT $=$ 2867 K, and $20\%$ brightness drop  compared with the input cold-white light, which is based on our simulation (red dashed curve). 
  The black long-dashed curve represents the reference light with CCT $=$ 2867 K.
  The QDCCF is prepared with emission wavelength/FWHM ($\lambda_\textnormal{R}$/$\Delta_\textnormal{R}$) 621/34 nm and film thickness 80 $\mu$m.
  (b) Spectra of the input day-white light (blue solid curve) and the measured (solid curves) as well as the simulated (dashed curved) warm-white light of lamp-2.
  (c) Time-dependent PL intensity of the QDCCF, where the red emission keeps initial intensity for over 1000 hours.
  The insets in (a) and (b) demonstrate the respective optimized devices.
  }\label{optimized-sample}
\end{figure}

Figure \ref{optimized-sample} (a) compares the measured and simulated spectra of the optimized warm-white light of lamp-1.
Besides the deviation of the yellow-phosphor emission in these spectra, the red emission in the measured spectrum is also weaker than that in the simulated one.
This is due to the lower quantum yield ($\eta_\textnormal{QY}\approx80\%$) of the QDs used for the QDCCF with $\lambda_\textnormal{R}=621$ nm, leading to that $\eta_\textnormal{EQE}$ decreases to a value of $60\%$.
If $\eta_\textnormal{QY}>90\%$, the resulting LE can be larger than 70 lm/W.
Nevertheless, this simple device has shown better performance and efficiency than that of most other warm-white light technologies and even most QD-LEDs \cite{Shirasaki&Bawendi&Bulovic:NPhoton.7.13(2013)(Emergence_CQD_light-emitting_tech)}, without modifying the chip efficiency, the phosphor composition, and the light-guide structure of the commercial lamp.
This indicates the importance of this optimization procedure in conjunction with the feasible tunability of QDs.

We also applied the QDCCFs to lamp-2 of day-white light (5792 K) with better colorimetric performance ($R_{a}=83.7$ and $R_{9}=14.8$).
Following the same optimization procedure, we have achieved warm-white light (3549 K) with $R_{a}=93.2$ and $R_{9}=55.3$, as shown in Fig.~\ref{optimized-sample}(b). 
With $\lambda_\textnormal{R}=617$ nm, the brightness drop reduces to 15\% due to less CCT difference between the day-white and warm-white light.
Furthermore, the colorimetric performance can be improved to $R_{a}=95.1$ and $R_{9}=78.6$ with $\lambda_\textnormal{R}=622$ nm and CCT$=3305$ K. 
Although the brightness drop of the sample of $R_{a}=95.1$ is 27\%, the reason is not only the larger emission wavelength, but also the lower $\eta_\textnormal{QY}$ of the corresponding QDs. 
Our simulation shows that lamp-2 has the potential to achieve $R_{a}=96$ and $R_{9}=96$ with limited brightness drop about $20\sim25\%$ as $\lambda_\textnormal{R}$ is shifted to 630 nm.

By exploiting the flat-lamp structure, our optimized devices also show excellent reliability, as shown in Fig.~\ref{optimized-sample}(b).
The PET-protected QDCCF was assembled into the flat-lamp module and tested with 4 W electrical power under ambient conditions over 1000 hours, where the photoluminescent (PL) intensity, CCT, and CRIs remain unchanged from the initial performance.
From a previous report \cite{Moon&Yoo:Opt.Mater.Express.4.2174(2014)(Optical_characteristics_longevity_QD-coated_WLED)}, without the PET protection, the emission of QDs quenches after 300 hours even for a remote-phosphor structure.
The operating temperature of QDCCF is about 45 $^{\circ}$C, slightly higher than the room temperature (30 $^{\circ}$C) due to down-conversion-induced heat. 
The successful fabrication of reliable high-CRI flat lamps with the robust QDCCFs is quite encouraging with regard to the development of next-generation lighting sources.

\subsection{Performance limits of QDCCFs}

\begin{table*}
  \caption{
  Optimization for given $\eta_\textnormal{EQE}$.
  }\label{Optimization for fixed EQE}
  \begin{tabular}{cccccccc}
    \hline
    $\eta_\textnormal{EQE}$ ($\%$) & $\lambda_\textnormal{R}$ (nm) & $\Delta_\textnormal{R}$ ($\geq$ 20 nm) & film thickness ($\mu$m) & LE ratio ($\%$) & CCT & $R_{a}$ & $R_{9}$ \\
    \hline
    50 & 619 & 20 & 81 & 85.8 & 3139 & 90 & 72 \\
    60 & 618 & 20 & 70 & 90.6 & 3197 & 90 & 68 \\
    70 & 617 & 20 & 63 & 94.1 & 3157 & 90 & 65 \\
    80 & 617 & 22 & 59 & 96.6 & 3141 & 90 & 67 \\
    90 & 617 & 29 & 57 & 98.8 & 3089 & 90 & 63 \\
    100 & 618 & 34 & 56 & 100.8 & 3030 & 90 & 60 \\
    \hline
  \end{tabular}
\end{table*}

We also estimated the limit of the device performance based on lamp-1, which could be further improved by tuning the FWHM ($\Delta_\textnormal{R}$) and external quantum efficiency ($\eta_\textnormal{EQE}$) of the QDCCFs.
For a critical situation, the minimum $\Delta_\textnormal{R}$ of contemporary QDs is about 20 nm \cite{Wang_et_al.:Nanoscale.7.2951(2015)(violet_ZnSe_QLED)}.
The advantage of small $\Delta_\textnormal{R}$ of QDs compared with that of traditional phosphors is to avoid energy waste in the deep-red region.
Meanwhile, the whole spectrum needs to be as continuous as possible to keep high color rendition.
Our algorithm provides a way to study this trade-off and suggests an optimal value of $\Delta_\textnormal{R}$.
The external quantum efficiency of a QDCCF is dominated by $\eta_\textnormal{QY}$ of the QDs and the design of film structure. 
Its improvement reduces the cost of QDs and matrix materials, and enhances the LE of the warm-white light devices.
Besides improving $\eta_\textnormal{QY}$, the light-extraction efficiency from guiding modes can be enhanced by modifying the surface structure such as diffuser attachment, microlens (or prism) structures, distributed Bragg reflectors, and photonic crystals \cite{Zhmakin:Phys.Rep.498.189(2011)(LED_light_extraction)}. 
However, these factors strongly depend on the synthesis process of QDs and manufacturability of QDCCFs.

Table \ref{Optimization for fixed EQE} shows the optimal $\lambda_\textnormal{R}$, $\Delta_\textnormal{R}$ (with minimum value of 20 nm), and film thickness for given $\eta_\textnormal{EQE}$, where $R_{a}$ is required to be 90 with largest LE ratio with respect to the input cold-white light.
If $\Delta_\textnormal{R}$ is smaller than that of our samples, the peak wavelength can be shorter than 621 nm, with thinner film thickness and larger LE.
For lower $\eta_\textnormal{EQE}$, the bandwidth is required to be as narrow as possible to achieve the specification with the largest LE.
If $\eta_\textnormal{EQE}$ is higher than 80\%, the optimal $\Delta_\textnormal{R}$ becomes larger than 20 nm, and the LE could be almost the same as the input cold-white light.

\section{Summary}

In summary, we developed a simulation algorithm for optimizing QD-based light-emitting devices.
We also practically fabricated quantum-dot color conversion films applied to two commercial white-light lamps for validation of the simulation. 
Under this simple flat-lamp structure, our devices show excellent reliability over 1000 hours and are expected to extend to various commercial products.
The first optimized device achieves warm-white light with $R_{a}=90.8$, $R_{9}=74.9$, and only 20\% brightness drop (68 lm/W) due to lowering the color temperature.
The second device can be designed either for excellent color rendition with $R_{a}=95.1$ and $R_{9}=78.6$ or for lower brightness drop about 15\% with $R_{a}=93.2$ and $R_{9}=55.3$.
The algorithm can be easily applied to other devices such as LEDs with remote phosphors or with on-chip phosphor package for desired CCT or individual CRIs ($R_{i}$, $i=1\sim15$). 
It is also useful for QD backlight units of liquid-crystal displays with given spectra of color filters, target white point, and color gamut.
Optimization essentially plays a significant role of QD-based technologies, which can highlight the feasible tunability of QDs for higher photometric and colorimetric performance. 
Thus our work paves the way for high-quality QD-based lighting and display devices.

\section{Experimental}

\subsection{Chemicals}

CdZnSeS/ZnS core/shell quantum dots (Taiwan Nanocrystals Inc.), sub-micrometer ZnO powder (99\%, Showa), methyl methacrylate (MMA, 99\%, Aldrich), LMA (96\%, Aldrich), acrylic acid (MA, 99\%, Aldrich), azobisisobutyronitrile (AIBN, 99\%, Aldrich) were used in an as-received condition without further purification.

\begin{figure}
  \includegraphics[width=0.5\linewidth]{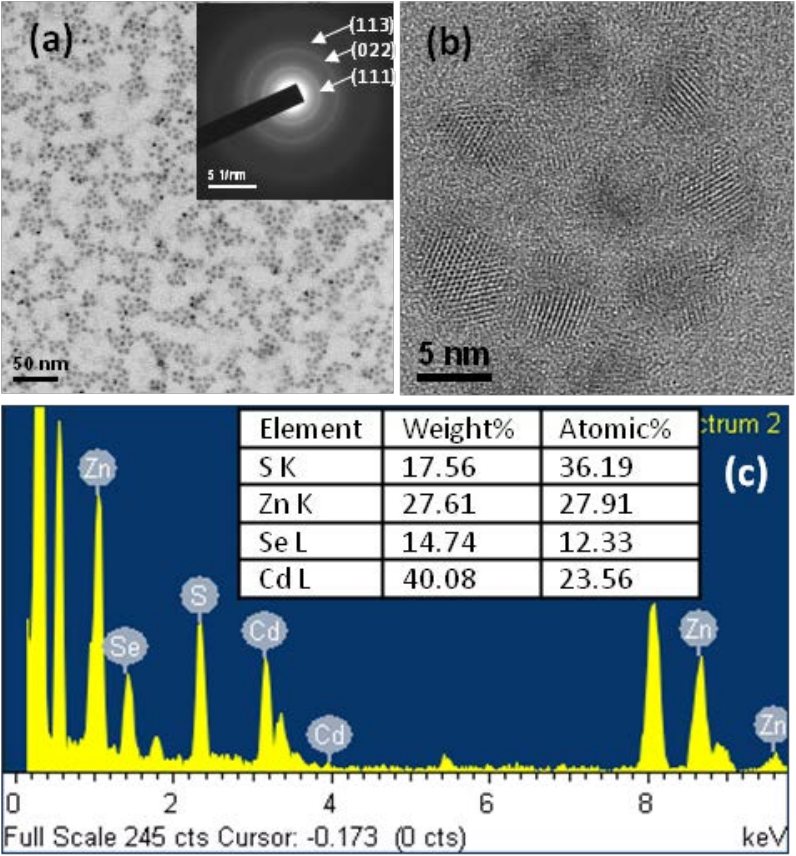}
  \caption{
  (a) TEM and (b) HRTEM micrographs of the CdZnSeS/ZnS QDs with emission wavelength of 609 nm. 
  The SAED pattern inserted in (a) shows a zinc-blende structure of the QDs. 
  (c) EDS patterns of the CdZnSeS/ZnS QDs.
  }\label{CQD_identification}
\end{figure}

The CdZnSeS/ZnS core/shell QDs were chosen with emission wavelength $605 \sim 615$ nm, FWHM 33 nm, and PL quantum yield about $80\sim90\%$. 
Figure \ref{CQD_identification}(a) shows the transmission-electron-microscopy (TEM) image of QDs with emission wavelength 609 nm and average size $6.5\pm0.3$ nm. 
The selected area electron diffraction (SAED) patterns [inserted in Fig.~\ref{CQD_identification}(a)] and high-resolution TEM [HRTEM, Fig.~\ref{CQD_identification}(b)] image reveal a high-quality zinc-blende (ZB) structure of the QDs.  
The atomic composition of the QDs is identified as 23.6\% Cd, 12.3\% Se, 27.9\% Zn, and 36.2\% S, from the energy-dispersive spectroscopy (EDS) shown in Fig.~\ref{CQD_identification}(c).

\subsection{Fabrication of QDCCF}

In the fabrication of QDCCFs, QDs (0.02 g), ZnO powder (0.02 g), PMMA$_{0.75}$-LMA$_{0.24}$-co-MA$_{0.01}$ (1.98 g), and toluene (7 mL) were mixed in a 20 mL flask by a magnetic stirrer for 1 day at room temperature. 
The QD-polymer-toluene solution (8 mL) was poured onto a PET-based release film. 
The QD films were formed by a blade coating machine and dried at room temperature for 1 hour and then at 60 $^\circ$C for 4 hours for removal of free toluene solvent. 
The thickness of the film can be controlled between 40 to 100 $\mu$m.
Finally, the dry film was encapsulated in the middle of two PET (20 $\mu$m) films by a laminating machine to avoid the contact with water and oxygen.


\begin{acknowledgement}


This work is supported by the Ministry of Science and Technology, Taiwan, Republic of China, Grant No. MOST 104-2119-M-007-014-MY2 and MOST 104-2113-M-269-001. 

\end{acknowledgement}




\bibliography{CQD}

\end{document}